\documentclass[conference]{IEEEtran}
\usepackage[final]{graphicx}
\usepackage{subfigure}
\usepackage{float}
\usepackage{amsmath}
\usepackage{cases}
\usepackage{color}
\usepackage{colortbl}
\usepackage{authblk}
\usepackage{algorithm}
\usepackage{algpseudocode}
\usepackage{amsmath}
\usepackage{amssymb}
\usepackage{booktabs}
\usepackage{setspace}
\usepackage{array}
\makeatletter
\renewcommand{\maketag@@@}[1]{\hbox{\m@th\normalsize\normalfont#1}}%
\makeatother
\usepackage{stfloats}
\usepackage{cite}
\usepackage{makecell}
\usepackage{multirow}

\def\BibTeX{{\rm B\kern-.05em{\sc i\kern-.025em b}\kern-.08em
    T\kern-.1667em\lower.7ex\hbox{E}\kern-.125emX}}
    
\ifCLASSINFOpdf
\else
\fi

\usepackage{caption}
\usepackage{mathtools}

\begin{document}
\title{Multipath Component-Aided Signal Processing for Integrated Sensing and Communication Systems}
\author[1]{Haotian Liu}
\author[1,*]{Zhiqing Wei}
\author[1]{Xiyang Wang} 
\author[1]{Yangyang Niu}
\author[1]{Yixin Zhang}
\author[1]{Huici Wu}
\author[1]{Zhiyong Feng}
\affil[1]{Beijing University of Posts and Telecommunications, Beijing 100876, China}
\affil[*]{Corresponding Author: Zhiqing Wei}
\affil[1]{Email:\{haotian\_liu, weizhiqing, wangxiyang, niuyy, yixin.zhang, dailywu, fengzy\} @bupt.edu.cn}

\maketitle

\begin{abstract} 
Integrated sensing and communication (ISAC) has emerged as a pivotal enabling technology for sixth-generation (6G) mobile communication system. The ISAC research in dense urban areas has been plaguing by severe multipath interference, propelling the thorough research of ISAC multipath interference elimination. However, transforming the multipath component (MPC) from enemy into friend is a viable and mutually beneficial option. In this paper, we preliminarily explore the MPC-aided ISAC signal processing and apply a space-time code to improve the ISAC performance. 
Specifically, we propose a symbol-level fusion for MPC-aided localization (SFMC) scheme to achieve robust and high-accuracy localization, and apply a Khatri-Rao space-time (KRST) code to improve the communication and sensing performance in rich multipath environment. 
Simulation results demonstrate that the proposed SFMC scheme has more robust localization performance with higher accuracy, compared with the existing state-of-the-art schemes. The proposed SFMC would benefit highly reliable communication and sub-meter level localization in rich multipath scenarios.
\end{abstract}

\IEEEpeerreviewmaketitle

\section{Introduction}
The envisioned next-generation wireless networks aim to endow mobile communication systems with integrated sensing and communication (ISAC) capabilities~\cite{Fanliu_2022,wei2024deep,cui2024}, enabling novel applications that rely heavily on high-speed communication and high-accuracy sensing capability. ISAC technique is appealing to both academia and industry and has the potential to transcend traditional communication paradigms.~\cite{cui2024,liu2023isac,Mohammad2023,wei2023carrier}.
By harnessing the multidimensional resource characteristics of mobile communication systems, the concept of multi-resource cooperative ISAC has been proposed~\cite{wei2024deep,wei2023carrier,liu2024carrier}, jointing space-time-frequency-code domain resources to obtain high-accuracy sensing and high-reliability communication performance.

However, mitigating the impact of multipath component (MPC) on high-accuracy sensing in dense urban areas is a formidable challenge~\cite{zhao_2020,laoudias2018survey}. 
Currently, many approaches focus on detecting and eliminating multipath effects~\cite{chen2002,lee2023seamless}.
Based on the concept of multi-resource cooperation, MPCs are anticipated to be utilized as spatial domain resources to further enhance sensing performance, which therefore would be helpful for ISAC systems~\cite{wei2024deep,Erik_2023}.

The research on MPC utilization based on communication signal focus on wireless localization.
Zhao \textit{et al.} in \cite{zhao2021} proposed a MPCs separation
method based on tensor decomposition, using the time of arrival (ToA) and angle of arrival (AoA) information of line-of-sight (LoS) path for localization. However, this method did not utilize the information from non-LoS (NLoS) paths. To this end, Gong \textit{et al.} in \cite{gong2022multipath} proposed an MPC-aided localization method. By establishing virtual anchors (VAs), the
joint localization of both LoS and NLoS paths information is achieved. This method fuses the target parameters carried by MPCs, which is a data-level fusion method of sensing information~\cite{liu2024carrier,wei2024deep}.

Overall, there is little work on MPC-aided ISAC signal processing. In existing work, the information of MPC has still not been fully utilized for MPC-aided localization scheme based on communication signal. To the best of our knowledge, this paper is the first attempt to study MPC-aided ISAC signal processing and the symbol-level fusion for MPC-aided localization (SFMC) scheme. Compared to existing data-level fusion schemes in \cite{zhao2021,gong2022multipath}, the proposed SFMC scheme can bring better sensing performance because it makes use of the distribution pattern of sensing information in complex space~\cite{liu2024carrier,wei2024deep,wei2023carrier}. At the same time, the Khatri-Rao space-time (KRST) code is applied to improve the performance of communication and sensing in rich multipath environment.
Simulation results demonstrate that the proposed SFMC scheme would benefit highly reliable communication and sub-meter level localization in rich multipath scenario.

The remainder of this paper is structured as follows. Section \ref{se2} outlines the system model. Section \ref{se3} presents a SFMC scheme. Section \ref{se4} details the simulation results, and Section \ref{se5} summarizes the paper.

\textit{Notations:} $\{\cdot\}$ stands for a set of various index values. $<\cdot>$ denotes extracting data from a coordinate. 
Black bold letters represent matrices or vectors.
$\mathbb{C}$ and $\mathbb{R}$ denote the set of complex and real numbers, respectively. 
$\left[\cdot\right]^{\text{T}}$, $\left[\cdot\right]^{*}$, $\text{diag}(\cdot)$, $\left[\!\left[ \cdot \right]\!\right]$,  $\equiv$, and $|\cdot|$ stand for the transpose operator, conjugate operator, diagonal operator, multilinear mappings, identically equal operator, and absolute operator, respectively. $\bullet $ and $\odot$ denote the outer product and Khatri–Rao product, respectively. $\left\|\cdot\right\|_{\text{F}}^2$ is the Frobenius norm. A complex Gaussian random variable $\mathbf{u}$ with mean $\mu_u$ and variance $\sigma_u^2$ is denoted by $\mathbf{u} \sim \mathcal{CN}\left(\mu_u,\sigma_u^2\right)$.

\section{System Model}\label{se2}
\begin{figure}
    \centering
    \includegraphics[width=0.40\textwidth]{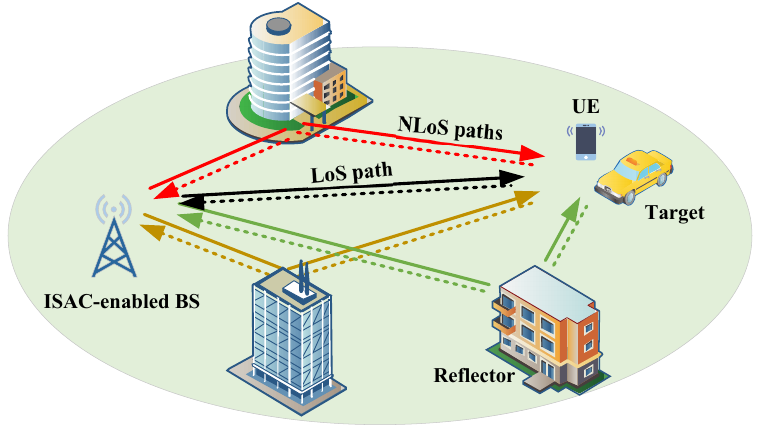}
    \caption{Downlink ISAC scenario for communicating with the UE and sensing the target, where MPC are utilized for signal processing at the BS.}
    \label{fig1}
\end{figure}

As shown in Fig. \ref{fig1}, we consider an ISAC-enabled base station (BS) for target sensing and downlink (DL) communication in rich multipath environment. The BS adopts a uniform linear array (ULA) with the antenna spacing being $d_r$, where $N_\text{T}$ transmit antennas and $N_\text{R}$ receive antennas are time-shared for communication and sensing~\cite{liu2024carrier}. 
For DL communication, the user equipment (UE) is equipped with $N_\text{U}$ receive antennas. For sensing,
we consider NLoS and LoS paths between BS and target. It should be noted that our work is also applicable when the LoS path does not exist.
Without loss of generality, 
we assume that the BS has prior physical knowledge of the  
$L$ surrounding reflectors, and that only the one-bounce specular reflection of the ISAC signal is considered~\cite{sen2010adaptive}.
The number of MPCs can be obtained by minimum description length (MDL) method~\cite{yokota2016robust}, which equals to the number of reflectors.

\subsection{STBC-MIMO-OFDM ISAC transmitter}
We consider a DL communication with $M$ code blocks, as well as sensing. $\mathbf{S}_n=\left[\mathbf{s}_n^1,\cdots,\mathbf{s}_n^m,\cdots,\mathbf{s}_n^M\right]\in\mathbb{C}^{N_\text{T}\times M}$ associated with the $n$-th subcarrier is the transmitted modulated symbol matrix, where the total number of subcarriers is $N_\text{c}$. The coding symbol matrix encoded by Khatri-Rao space-time (KRST) coding scheme~\cite{Sidiropoulos2002} is $\text{diag}\left(\Xi\mathbf{s}_n^m\right)\mathbf{C}_0^\text{T}$, where $\Xi=\frac{1}{\sqrt{N_\text{T}}}\tilde{\boldsymbol{F}}_{n}\text{diag}\left\{1,e^{j\pi/2N_\text{T}},\cdots,e^{j\pi(N_\text{T}-1)/2N_\text{T}}\right\}\in\mathbb{C}^{N_\text{T} \times N_\text{T}}$ is constellation rotation matrix with $\tilde{\boldsymbol{F}}_{n}\in\mathbb{C}^{N_\text{T} \times N_\text{T}}$ being a inverse discrete Fourier transform (IDFT) matrix; $\mathbf{C}_0\in\mathbb{C}^{K\times N_\text{T}}$ is a scaled semi-unitary matrix with $\mathbf{C}_0^\text{H}\mathbf{C}_0=\mathbf{I}_{N_\text{T}}\left(K\le N_\text{T}\right)$ and $K$ being the code length.

\subsection{Communication Signal Model}
In rich multipath environment, the KRST coding is introduced to combat multipath fading while enabling flexible adjustment between the achievable rate and the diversity gain.~\cite{Sidiropoulos2002}.

\subsubsection{Channel model}
Since the LoS paths exist between the BS and the UE, we assume that the channel matrix $\mathbf{H}_C\in\mathbb{C}^{N_\text{U}\times N_\text{T}}$ is uncorrelated Rician fading channel and its elements can be expressed as~\cite{sanguinetti2018theoretical}
\begin{equation}\label{eq1} \left[\mathbf{H}_C\right]_{u,p}=\sqrt{\beta\frac{\kappa}{1+\kappa}}h_{u,p}^{Ri}+\sqrt{\beta\frac{1}{1+\kappa}}h_{u,p}^{Ra},
\end{equation}
where $u\in\{0,1,\cdots,N_\text{U}-1\}$ and $p\in\{0,1,\cdots,N_\text{T}-1\}$ are the indices of receive UE antennas and transmit BS antennas, respectively; $\beta$ is path loss and $\kappa \ge 0$ is the Rician factor; $h_{u,p}^{Ri}$ is a deterministic
vector that associates with the LoS path, and $h_{u,p}^{Ra}\sim\mathcal{CN}\left(0,1
\right)$~\cite{sanguinetti2018theoretical}. 
\subsubsection{Received signal}
If the channel is stationary for the duration of code length $K$, the received signal at the UE in the $n$-th subcarrier during the $m$-th KRST code block is denoted by
\begin{equation} \label{eq2}
\mathbf{Y}_{n,m}^C=\mathbf{H}_C\text{diag}\left(\Xi\mathbf{s}_n^m\right)\mathbf{C}_0^\text{T}+\mathbf{Z}_{n,m}^C,
\end{equation}
where $\mathbf{Y}_{n,m}^C\in\mathbb{C}^{N_\text{U}\times K}$ and $\mathbf{Z}_{n,m}^C\in\mathbb{C}^{N_\text{U}\times K}$ is a additive white Gaussian nosie (AWGN) matrix. The KRST codes have the capability of blind channel state information (CSI) for communication data recovery, which can be achieved by the blind-KRST decoding technique. Due to space constraints, readers are referred to \cite{Sidiropoulos2002} for details of the blind-KRST decoding technique. With the KRST code, the rate of transmission is $\left(\frac{N_\text{T}}{K}\right)\log_2\left(\varphi \right)$ bit / (code length), where $\varphi$ is the order of modulation.

\subsection{Sensing Signal Model}
For the sensing in rich multipath environment, a SFMC scheme is proposed to achieve robust, high-accuracy target localization.

In the sensing duration of $M$ KRST code blocks, for a far-field point target with coordinate $\left(x_\text{ta},y_\text{ta}\right)$ moving with a absolute velocity $\Vec{v}$, the received ISAC signal in BS side of the $n$-th subcarrier during the $m$-th KRST code block is~\cite{liu2024carrier}
\begin{equation}  \label{eq3}
\mathbf{Y}_{n,m}^S=\sum_{l=0}^L\left[
\begin{array}{l}
 \sqrt{P_\text{t}^l}\alpha_l e^{j2\pi f_{\text{D},l} mT}e^{-j2\pi n\Delta f \tau_l}  \\
\times\textbf{a}_\text{r}\left(\theta_l\right)\textbf{a}_\text{t}^\text{T}\left(\theta_l\right)\text{diag}\left(\Xi\mathbf{s}_n^m\right)\mathbf{C}_0^\text{T}
\end{array}\right]+ \mathbf{Z}_{n,m}^S,
\end{equation}
where $\mathbf{Y}_{n,m}^S\in \mathbb{C}^{N_\text{R}\times K}$, $l\in\{0,1,\cdots, L\}$ denotes the index of MPCs; $l=0$ refers to the LoS path; $\alpha_l$ and $P_\text{t}^l$ are the attenuation and transmit power of the $l$-th MPC, respectively; $\sum_{l=0}^{L}P_\text{t}^l$ is the total transmit power;
$f_{\text{D},l}=\frac{2v_lf_\text{c}}{c_0}$ is the Doppler frequency shift, where $v_l$ is the relative velocity with respect to the $l$-th reflector and $c_0$ is the speed of light;
$\tau_l=\frac{2R_l}{c_0}$ is the delay with $R_l$ being the one-way distance of the $l$-th MPC; $T$ is total OFDM symbol duration;
$\theta_l$ is AoA of the $l$-th MPC and
$\mathbf{Z}_{n,m}^S \sim \mathcal{CN}(0,\sigma^2)$ is an AWGN matrix;
$\textbf{a}_\text{r}(\cdot)$ and $\textbf{a}_\text{t}(\cdot)$ are the receive and transmit steering vectors, expressed in (\ref{eq4}) and (\ref{eq5}), respectively~\cite{liu2024carrier}.
\begin{equation}\label{eq4}
    \textbf{a}_\text{r}(\cdot)=\left[1, \cdots, e^{j2\pi\frac{d_r}{\lambda} k\sin(\cdot)}, \cdots, e^{j2\pi \frac{d_r}{\lambda}(N_\text{R}-1)\sin(\cdot)}\right]^\text{T},
\end{equation}
\begin{equation}\label{eq5}
\textbf{a}_\text{t}(\cdot)=\left[1, \cdots, e^{j2\pi\frac{d_r}{\lambda} p\sin(\cdot)}, \cdots, e^{j2\pi\frac{d_r}{\lambda} (N_\text{T}-1)\sin(\cdot)}\right]^\text{T},
\end{equation}
where $\lambda$ is wavelength and $k\in\{0, 1, \cdots, N_\text{R}-1\}$ is the index of receive BS antenna. 

\section{Symbol-Level Fusion for MPC-Aided Localization scheme}\label{se3}
\begin{figure*}
    \centering
    \includegraphics[width=0.85\textwidth]{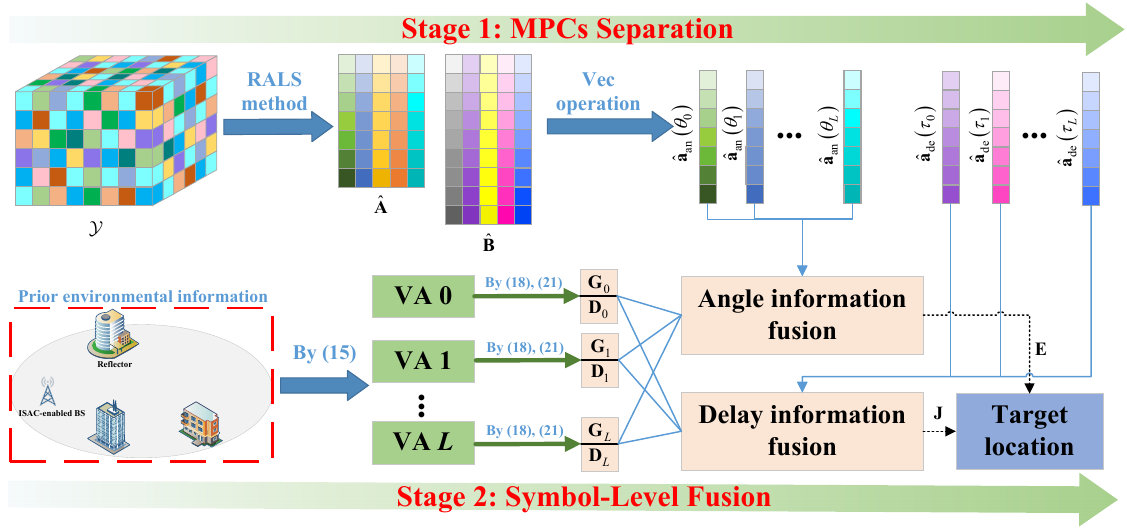}
    \caption{Illustration of the proposed SFMC scheme}
    \label{fig2}
\end{figure*}

As shown in Fig.~\ref{fig2}, the proposed SFMC scheme involves two stages: MPCs separation stage and symbol-level fusion stage.  

\subsection{MPCs Separation Stage}\label{sec3-a}
In this stage, the mixed MPCs signal is formed as a three-order tensor and is separated by the regularized
alternating least squares (RALS) method~\cite{li2013some}. 
\subsubsection{Tensor-based received ISAC signal model}
After the known code symbol matrix 
$\text{diag}\left(\Xi\mathbf{s}_n^m\right)\mathbf{C}_0^\text{T}$ in $\mathbf{Y}_{n,m}^S$ is cancelled, we can see that the sensing information in each slot of the $m$-th code block is identical, which can be coherently accumulated to obtain coding gain. Then, the third-order tensor form of the received ISAC signal in $N_\text{c}$ subcarriers during $M$ code block can be expressed as~\cite{gong2022multipath}
\begin{equation}\label{eq6}
\begin{aligned}
     \mathcal{Y}&\approx \left[\!\left[ \mathbf{A},\mathbf{B},\mathbf{C} \right]\!\right] + \mathcal{Z}\\& \equiv \sum_{l=0}^{L}\mathbf{a}_{\text{an}}\left(\theta_l\right)\bullet  \mathbf{a}_{\text{de}}\left(\tau_l\right)\bullet   \mathbf{f}\left(f_{\text{D},l}\right)+ \mathcal{Z},
\end{aligned}
\end{equation}
where $\mathcal{Y} \in \mathbb{C}^{N_\text{R}\times N_\text{c}\times M}$, and $\mathcal{Z}$ is an AWGN tensor with the element $\mathcal{Z}(k,n,m)$ following $\mathcal{CN}(0, \sigma_S^2)$; The factor matrices $\mathbf{A}=\left[\mathbf{a}_{\text{an}}\left(\theta_0\right),\mathbf{a}_{\text{an}}\left(\theta_1\right),\cdots, \mathbf{a}_{\text{an}}\left(\theta_L\right)\right]\in \mathbb{C}^{N_\text{R}\times(L+1)}$, $\mathbf{B}=\left[\mathbf{a}_{\text{de}}\left(\tau_0\right),\mathbf{a}_{\text{de}}\left(\tau_1\right),\cdots, \mathbf{a}_{\text{de}}\left(\tau_L\right)\right]\in \mathbb{C}^{N_\text{c}\times(L+1)}$, and $\mathbf{C}=\left[\mathbf{f}\left(f_{\text{D},0}\right),\mathbf{f}\left(f_{\text{D},1}\right),\cdots, \mathbf{f}\left(f_{\text{D},L}\right)\right]\in \mathbb{C}^{M\times(L+1)}$; $\mathbf{a}_{\text{an}}\left(\theta_l\right)$, $\mathbf{a}_{\text{de}}\left(\tau_l\right)$, and $\mathbf{f}\left(f_{\text{D},l}\right)$ are the vectors of the $l$-th MPC for the $1$-st, $2$-nd, and $3$-rd dimension, expressed in (\ref{eq7}), (\ref{eq8}), and (\ref{eq9}), respectively.
\begin{equation}\label{eq7}
\mathbf{a}_{\text{an}}\left(\theta_l\right)=\textbf{a}_\text{r}\left(\theta_l\right)\textbf{a}_\text{t}^\text{T}\left(\theta_l\right)\mathbf{e}, 
\end{equation}
\begin{equation}\label{eq8}
\mathbf{a}_{\text{de}}\left(\tau_l\right)=\left[1, e^{-j2\pi\Delta f\tau_l}, \cdots, e^{-j2\pi(N_\text{c}-1)\Delta f\tau_l}\right]^\text{T}, 
\end{equation}
\begin{equation}\label{eq9}
 \mathbf{f}\left(f_{\text{D},l}\right)= \left[1, e^{j2\pi f_{\text{D},l}T}, \cdots, e^{j2\pi f_{\text{D},l}(M-1)}\right]^\text{T},  
\end{equation}
where $\mathbf{e}\in\mathbb{R}^{N_\text{T}\times 1}$ is a unit vector.

Then, the three metricized versions of (\ref{eq6}) are expressed in (\ref{eq10}), (\ref{eq11}), and (\ref{eq12}), respectively~\cite{kolda2009tensor}.
\begin{equation}\label{eq10}
\mathbf{Y}_{(1)}\approx\mathbf{A}\left(\mathbf{C}\odot \mathbf{B}\right)^{\text{T}}\in\mathbb{C}^{N_\text{R}\times N_\text{c}M}, 
\end{equation}
\begin{equation}\label{eq11}
\mathbf{Y}_{(2)}\approx\mathbf{B}\left(\mathbf{C}\odot \mathbf{A}\right)^{\text{T}}\in\mathbb{C}^{N_\text{c}\times N_\text{R}M},     
\end{equation}
\begin{equation}\label{eq12}
\mathbf{Y}_{(3)}\approx\mathbf{C}\left(\mathbf{B}\odot \mathbf{A}\right)^{\text{T}}\in\mathbb{C}^{M\times N_\text{R}N_\text{c}}.   
\end{equation}

\subsubsection{CANDECOMP/PARAFAC (CP) decomposition}
For (\ref{eq6}), the CP decomposition problem is transformed into a low-rank approximation problem, which is represented as
\begin{equation}\label{eq13}
    \min_{\mathbf{\hat{A}}, \mathbf{\hat{B}}, \mathbf{\hat{C}}}\left\|\mathcal{Y}-\left[\!\left[ \mathbf{\hat{A}}, \mathbf{\hat{B}}, \mathbf{\hat{C}} \right]\!\right]\right\|_{\text{F}}^2.
\end{equation}
The solution to (\ref{eq13}) typically employs the ALS method~\cite{kolda2009tensor}. 
Considering the slow convergence and local optimization of ALS method, this paper adopts an RALS method~\cite{li2013some} to obtain the high convergence speed and precision. 

Similar to ALS, RALS recasts the (\ref{eq13}) to three coupled RALS subproblems, which are expressed as
\begin{equation}\label{eq14}
{\fontsize{8.5}{10} \begin{aligned}
\mathbf{A}^{i+1}&=\underset{\hat{\mathbf{A}}}{\text{argmin}}\|\mathbf{Y_{(1)}}-\hat{\mathbf{A}}(\mathbf{C}^i\odot\mathbf{B}^i)^\text{T}\|_\text{F}^2+\lambda\|\mathbf{A}^i-\hat{\mathbf{A}}\|_\text{F}^2, \\
\mathbf{B}^{i+1}&=\underset{\hat{\mathbf{B}}}{\text{argmin}}\|\mathbf{Y_{(2)}}-\hat{\mathbf{B}}(\mathbf{C}^i\odot\mathbf{A}^{i+1})^\text{T}\|_\text{F}^2+\lambda\|\mathbf{B}^i-\hat{\mathbf{B}}\|_\text{F}^2, \\
\mathbf{C}^{i+1}&=\underset{\hat{\mathbf{C}}}{\text{argmin}}\|\mathbf{Y_{(3)}}-\hat{\mathbf{C}}(\mathbf{B}^{i+1}\odot\mathbf{A}^{i+1})^\text{T}\|_\text{F}^2+\lambda\|\mathbf{C}^i-\hat{\mathbf{C}}\|_\text{F}^2,
\end{aligned} }
\end{equation}
where $\lambda>0$ is the regularization hyperparameter~\cite{li2013some};
$\hat{\mathbf{A}}$, $\hat{\mathbf{B}}$, and $\hat{\mathbf{C}}$ denote the estimated factor matrices that minimize the function value in (\ref{eq14}).
The obtained outputs $\hat{\mathbf{A}}=\left[\hat{\mathbf{a}}_{\text{an}}\left(\theta_0\right),\hat{\mathbf{a}}_{\text{an}}\left(\theta_1\right),\cdots, \hat{\mathbf{a}}_{\text{an}}\left(\theta_L\right)\right]$ and $\hat{\mathbf{B}}=\left[\hat{\mathbf{a}}_{\text{de}}\left(\tau_0\right),\hat{\mathbf{a}}_{\text{de}}\left(\tau_1\right),\cdots, \hat{\mathbf{a}}_{\text{de}}\left(\tau_L\right)\right]$ can be used to estimate target's location based on symbol-level fusion. 

\subsection{Symbol-Level Fusion Stage}\label{sec4-B}
In the symbol-level fusion stage, the virtual anchors (VAs) are constructed based on the prior environment information, and the $\hat{\mathbf{A}}$ and $\hat{\mathbf{B}}$ are performed symbol-level fusion to obtain the target's location. 

 We define the direction and center coordinate of the $l'\in\{1,2,\cdots, L\}$-th reflector as $\phi_{l'}\in\left[0, \pi\right]$ and $(x_{l'}, y_{l'})$, respectively. Therefore, the coordinate of the ${l'}$-th VA is
\begin{equation}\label{eq15}
 \begin{aligned}
    x_{{l'},\text{VA}}&=\frac{-2\left[y_{l'}-\tan (\phi_{l'}) \cdot x_{l'}\right] \tan (\phi_{l'})}{1+\tan ^{2}(\phi_{l'})}, \\
    y_{{l'},\text{VA}}&=\frac{2\left(y_{l'}-\tan (\phi_{l'}) \cdot x_{l'}\right)}{1+\tan ^{2}(\phi_{l'})}.
\end{aligned}   
\end{equation}
For $l'$-th VA, the column vectors associated with it are $\hat{\mathbf{a}}_\text{de}\left(\tau_{l'}\right)$ and $\hat{\mathbf{a}}_\text{an}\left(\theta_{l'}\right)$.
It is noted that the $0$-th VA  is regarded as the BS with the coordinate being $(x_{{0},\text{VA}}, y_{{0},\text{VA}})$.

\subsubsection{Angle information fusion}
The approach to angle information fusion involves creating an angle matching matrix according to the candidate coordinates of the target and then fusing it with the angles information $\{\hat{\mathbf{a}}_\text{an}\left(\theta_l\right)\}_{l=0:L}$ obtained in the Section~\ref{sec3-a} to screen out the maximum likelihood coordinate. The detailed procedures are as follows.

\textit{Step 1}: An interesting detection scope of length and width $\Omega\Delta R$ is gridded with grid size $\Delta R$ to obtain a searching matrix $\mathbf{P}\in\mathbb{C}^{\Omega\times\Omega}$, where $\left[\mathbf{P}\right]_{\rho ,\varsigma}=\left(x_{\rho ,\varsigma}^\text{g},y_{\rho ,\varsigma}^\text{g}\right)$ is the candidate coordinate with $\rho ,\varsigma \in \{1,2,\cdots,\Omega\}$.

\textit{Step 2}: According to the coordinates of VAs and candidate grid points, the angle candidate vectors between VAs and candidate grid points are obtained, and the $l$-th angle candidate vector associated with the $l$-th VA is $\vec{\mathbf{G}}_l\in\mathbb{C}^{1\times\Omega^2}$, where the $\xi$-th element of $\vec{\mathbf{G}}_l$ is 
\begin{equation}\label{eq16}
\vec{\mathbf{G}}_l(\xi)=\arctan{\frac{\left(y_{l,\text{VA}}-\left[\text{vec}(\mathbf{P})\right]_\xi^\text{T}\{2\}\right)}{\left(x_{l,\text{VA}}-\left[\text{vec}(\mathbf{P})\right]_\xi^\text{T}\{1\}\right)}},  
\end{equation}
and $\xi\in\{1,2,\cdots,\Omega^2\}$.
We notice that the angle $\hat{\mathbf{a}}_\text{an}(\theta_l)$ is not the angle between the $l$-th VA and far-field target. Therefore, the angle candidate vectors need to be transformed.

\textit{Step 3}: For the $l$-th angle candidate vector $\vec{\mathbf{G}}_l$, based on the geometric relationship, we can obtain the transformed angle candidate vector $\vec{\mathbf{G}}_l^-\in\mathbb{C}^{1\times\Omega^2}$, expressed as 
\begin{equation}\label{eq17}
  \vec{\mathbf{G}}_l^- = \left\{\begin{matrix}
 |\vec{\mathbf{G}}_l|\text{diag}\left[\mathbf{s}(-2\phi_{l}+2\pi)\right], & y_{l,\text{VA}}>0\\
 - |\vec{\mathbf{G}}_l|\text{diag}\left[\mathbf{s}(2\phi_{l}-2\pi)\right],& y_{l,\text{VA}}<0
\end{matrix}\right.
\end{equation}
where $\mathbf{s}\in\mathbb{R}^{1\times \Omega^2}$ is a unit vector. It is noted that the $0$-th transformed angle candidate vector associated with the BS is $\vec{\mathbf{G}}_0^-=\vec{\mathbf{G}}_0$.

\textit{Step 4}: Based on the transformed angle candidate vectors and (\ref{eq4}), we can obtain the angle matching matrices $\{\mathbf{G}_l\}_{l=0:L}$ and the $l$-th angle matching matrix $\mathbf{G}_l\in\mathbb{C}^{N_\text{R}\times\Omega^2}$ is expressed as
{\fontsize{7}{7} \begin{equation}\label{eq18}
 \mathbf{G}_l=  \begin{bmatrix}
 1 & \cdots & 1 \\
 e^{j2\pi\frac{d_r}{\lambda}\sin\left(\left[\vec{\mathbf{G}}_l^-\right]_1\right)} & \cdots & e^{j2\pi\frac{d_r}{\lambda}\sin\left(\left[\vec{\mathbf{G}}_l^-\right]_{\Omega^2}\right)} \\
 \vdots &  \ddots & \vdots \\
 e^{j2\pi\frac{d_r}{\lambda}(N_\text{R}-1)\sin\left(\left[\vec{\mathbf{G}}_l^-\right]_1\right)} & \cdots & e^{j2\pi\frac{d_r}{\lambda}(N_\text{R}-1)\sin\left(\left[\vec{\mathbf{G}}_l^-\right]_{\Omega^2}\right)}
\end{bmatrix}.
\end{equation} }

\textit{Step 5}: Finally, $L+1$ angle matching matrices $\{\mathbf{G}_l\}_{l=0:L}$ and corresponding $L+1$ angle vectors $\{\hat{\mathbf{a}}_\text{an}\left(\theta_l\right)\}_{l=0:L}$ are fused to obtain a angle fusion vector $\mathbf{E}\in\mathbb{C}^{1\times\Omega^2}=\frac{1}{\sqrt{L+1}}\sum_{l=0}^{L}|\mathbf{E}_l|$, where $\mathbf{E}_l$ is the $l$-th angle profile, expressed as
\begin{equation}\label{eq19}
\mathbf{E}_l=\left[\hat{\mathbf{a}}_\text{an}\left(\theta_l\right)\right]^\text{T}\mathbf{G}_l^*.
\end{equation}

So far, the obtained angle fusion vector can be used to achieve target localization. In order to further improve the localization accuracy, delay information can be combined for fusion. The procedure is similar to the procedure of angle information fusion. Due to space reasons, a brief description is given as follows.

\subsubsection{Delay information fusion}
According to $\mathbf{P}$, we obtain $L+1$ delay candidate vectors, and the $l$-th delay candidate vector is $\vec{\mathbf{D}}_l\in\mathbb{C}^{1\times\Omega^2}$, where the $\xi$-th element of $\vec{\mathbf{D}}_l$ is 
\begin{equation}\label{eq20}
\scalebox{1}{$
\vec{\mathbf{D}}_l\left(\xi\right)=\frac{2\sqrt{\left(y_{l,\text{VA}}-\left[\text{vec}(\mathbf{P})\right]_\xi^\text{T}\{2\}\right)^2+\left(x_{l,\text{VA}}-\left[\text{vec}(\mathbf{P})\right]_\xi^\text{T}\{1\}\right)^2}}{c_0}.
$}
\end{equation}
Then, based on the delay candidate vectors and (\ref{eq8}), the $L+1$ delay matching matrices are obtained and the $l$-th delay matching matrix $\mathbf{D}_l\in\mathbb{C}^{N_\text{c}\times \Omega^2}$ is expressed as
\begin{equation}\label{eq21}
\scalebox{0.75}{$
  \mathbf{D}_l= \begin{bmatrix}
 1 & 1 & \cdots & 1 \\
 e^{-j2\pi\Delta f\left[\vec{\mathbf{D}}_l\right]_1} & e^{-j2\pi\Delta f\left[\vec{\mathbf{D}}_l\right]_2} & \cdots & e^{-j2\pi\Delta f\left[\vec{\mathbf{D}}_l\right]_{\Omega^2}}\\
 \vdots & \vdots & \ddots & \vdots \\
 e^{-j2\pi\Delta f\left(N_\text{c}-1\right)\left[\vec{\mathbf{D}}_l\right]_1} & e^{-j2\pi\Delta f\left(N_\text{c}-1\right)\left[\vec{\mathbf{D}}_l\right]_2} & \cdots & e^{-j2\pi\Delta f\left(N_\text{c}-1\right)\left[\vec{\mathbf{D}}_l\right]_{\Omega^2}}
\end{bmatrix}.$}
\end{equation}

Finally, $L+1$ delay matching matrices $\{\mathbf{D}_l\}_{l=0:L}$ and corresponding $L+1$ delay vectors $\{\hat{\mathbf{a}}_\text{de}\left(\tau_l\right)\}_{l=0:L}$ are fused to obtain a delay fusion vector $\mathbf{J}\in\mathbb{C}^{1\times\Omega^2}=\frac{1}{\sqrt{L+1}}\sum_{l=0}^{L}|\mathbf{J}_l|$, where $\mathbf{J}_l$ is the $l$-th delay profile, expressed as
\begin{equation}\label{eq22}
\mathbf{J}_l=\left[\hat{\mathbf{a}}_\text{de}\left(\tau_l\right)\right]^\text{T}\mathbf{D}_l^*.
\end{equation}

To obtain the target's location, the peak value of $\mathbf{J}+\mathbf{E}$ is searched to obtain the peak index value $\hat{\xi}$. The value of $\left[\text{vec}(\mathbf{P})\right]_{\hat{\xi}}^\text{T}$ is the estimated target's location, denoted by $\left(\hat{x}_\text{ta},\hat{y}_\text{ta}\right)$, and the \textbf{Algorithm 1} demonstrates the proposed SFMC scheme. {\color{blue}It should be noted that the algorithm presented in this paper is currently applicable to stationary ISAC systems. For mobile ISAC systems, an environmental map needs to be constructed prior to the implementation of the SFMC scheme to ensure the preservation of the prior geometric relationship between the ISAC systems and the reflectors. Additionally, the potential of the temporary MPC remains to be further explored, which will be the focus of our future work.}

\begin{table}[!ht]
\centering
\label{tab2}
\resizebox{0.90\linewidth}{!}{
\setlength{\arrayrulewidth}{1.5pt}
\begin{tabular}{rllll}
\hline
\multicolumn{5}{l}{\textbf{Algorithm 1:} The Proposed SFMC Scheme}   \\ \hline
\multirow{-4}{*}{\textbf{Input:} }               & \multicolumn{4}{l}{\begin{tabular}[c]{@{}l@{}}The received three-order tensor $\mathcal{Y}$ in (\ref{eq6});\\ The directions and center coordinates of reflectors \\ $\{\left(x_{l'},y_{l'}\right)\}_{l'=1:L}$ and $\{\phi_{l'}\}_{l'=1:L}$;\\ The grid size $\Delta R$. \end{tabular}} \\
\textbf{Output:}               & \multicolumn{4}{l}{The target's location result $\left(\hat{x}_\text{ta},\hat{y}_\text{ta}\right)$.} \\ 
\multicolumn{5}{l}{\textbf{MPCs Separation Stage:}} \\
\multirow{-2}{*}{1:}       & \multicolumn{4}{l}{\begin{tabular}[c]{@{}l@{}}Obtain the factor matrices $\hat{\mathbf{A}}$ and $\hat{\mathbf{B}}$ by (\ref{eq13}),\\  (\ref{eq14}), and RALS method;\end{tabular}} \\
2:        & \multicolumn{4}{l}{Output $\hat{\mathbf{A}}=\left\{\hat{\mathbf{a}}_\text{an}\left(\theta_0\right),\hat{\mathbf{a}}_\text{an}\left(\theta_1\right),\cdots,\hat{\mathbf{a}}_\text{an}\left(\theta_L\right)\right\}$;}  \\
3:        & \multicolumn{4}{l}{Output $\hat{\mathbf{B}}=\left\{\hat{\mathbf{a}}_\text{de}\left(\tau_0\right),\hat{\mathbf{a}}_\text{de}\left(\tau_1\right),\cdots,\hat{\mathbf{a}}_\text{de}\left(\tau_L\right)\right\}$;}  \\
\multicolumn{5}{l}{\textbf{Symbol-Level Fusion Stage:}} \\
\multirow{-2}{*}{4:}     & \multicolumn{4}{l}{\begin{tabular}[c]{@{}l@{}}Obtain the coordinates of VAs $\{(x_{l',\text{VA}},y_{l',\text{VA}})\}_{l'=1:L}$ by\\ $\{\left(x_{l'},y_{l'}\right)\}_{l'=1:L}$, $\{\phi_{l'}\}_{l'=1:L}$, and (\ref{eq15});\end{tabular}}  \\
\multirow{-2}{*}{5:}                              & \multicolumn{4}{l}{\begin{tabular}[c]{@{}l@{}}Obtain searching matrix $\mathbf{P
}$ by $\Delta R$ and \\ obtain $\{\vec{\mathbf{G}}_l\}_{l=0:L}$ by $\mathbf{P
}$ and (\ref{eq16});\end{tabular}}  \\
\multirow{-2}{*}{6:}                              & \multicolumn{4}{l}{\begin{tabular}[c]{@{}l@{}}Obtain transformed angle candidate vector $\{\vec{\mathbf{G}}_l^-\}_{l=0:L}$ by \\ $\{\vec{\mathbf{G}}_l\}_{l=0:L}$ and (\ref{eq17});\end{tabular}}  \\
\multirow{-2}{*}{7:}                            & \multicolumn{4}{l}{\begin{tabular}[c]{@{}l@{}}Obtain the angle matching matrices $\{\mathbf{G}\}_{l=0:L}$ and \\ angle fusion vector $\mathbf{E}$ by (\ref{eq18}) and (\ref{eq19});\end{tabular}}   \\
8:                              & \multicolumn{4}{l}{Obtain delay candidate vectors $\{\vec{\mathbf{D}}_l\}_{l=0:L}$ by $\mathbf{P}$ and (\ref{eq20});}  \\
\multirow{-2}{*}{9:}                            & \multicolumn{4}{l}{\begin{tabular}[c]{@{}l@{}}Obtain the delay matching matrices $\{\mathbf{D}\}_{l=0:L}$ and \\ delay fusion vector $\mathbf{J}$ by (\ref{eq21}) and (\ref{eq22});\end{tabular}}   \\
10:                            & \multicolumn{4}{l}{Obtain the peak index $\hat{\xi}$ by $\mathbf{J}+\mathbf{E}$;}  \\
11:                            & \multicolumn{4}{l}{Obtain the estimated target's location $\left(\hat{x}_\text{ta},\hat{y}_\text{ta}\right)$.}    \\
\hline
\end{tabular}}
\end{table}

\begin{table*}
	\caption{Simulation parameters \cite{wei2023carrier,liu2024carrier}}
	\label{tab_simulation}
	\renewcommand{\arraystretch}{1.0} 
	\begin{center}\resizebox{0.8\linewidth}{!}{
		\begin{tabular}{|m{0.1\textwidth}<{\centering}| m{0.3\textwidth}<{\centering}| m{0.13\textwidth}<{\centering}| m{0.1\textwidth}<{\centering}|m{0.3\textwidth}<{\centering}| m{0.15\textwidth}<{\centering}|}
			\hline
			\textbf{Symbol} & \textbf{Parameter} & \textbf{Value} & \textbf{Symbol} & \textbf{Parameter} & \textbf{Value} \\
			\hline
			$N_\text{c}$	& Number of subcarriers in BS & 1024 & $M_{\text{sym}}$	& Number of code blocks & 14  \\
                \hline
			$N_\text{R}$, $N_\text{T}$ & Number of receive and transmit antennas in BS & 128, 128 & $L$& Number of MPCs & 1-4 \\
			\hline
			$f_\text{c}$& Carrier frequency & 3.5 GHz & $\Delta f$  & Subcarrier spacing  & 30 kHz \\
			\hline
		 $(x,y)$ & Coordinate of the ISAC BS  & (0, 0) m	& $(x_1,y_1)$, $\phi_1$	& The center coordinate and direction of the $1$-st reflector & (60, 70) m, $0^\circ $  \\			
			\hline
		  $(x_2,y_2)$, $\phi_2$ & The center coordinate and direction of the $2$-nd reflector & (40, 100) m, $0^\circ $ & $(x_3,y_3)$, $\phi_3$	& The center coordinate and direction of the $3$-rd reflector & (50, -100) m, $45^\circ$  \\
                \hline
          $(x_\text{tr},y_\text{tr})$& The coordinate of target & (100, 50) m & $|\Vec{v}|$, $\theta_\text{v}$	& The magnitude and angle of target's velocity & 100 m/s, $27.6923^\circ$  \\
			\hline
		\end{tabular}}
	\end{center}
\end{table*}

\section{Simulation Results}\label{se4}
In this section, the feasibility and superiority of localization of the proposed SFMC scheme is validated through simulation results. The simulation parameters are listed in Table~\ref{tab_simulation} and the simulation results are obtained via $10^4$ times of Monte Carlo simulations.

\subsection{Sensing Performance}\label{se4-A}
In this section, the root mean square error (RMSE) is applied to verify the feasibility and superiority of the proposed SFMC scheme in sensing performance. {\color{blue}}
\subsubsection{RMSE of localization}
\begin{figure*}[!htbp]
	\centering
	\subfigure[One MPC] {\label{fig4.a}\includegraphics[width=.24\textwidth]{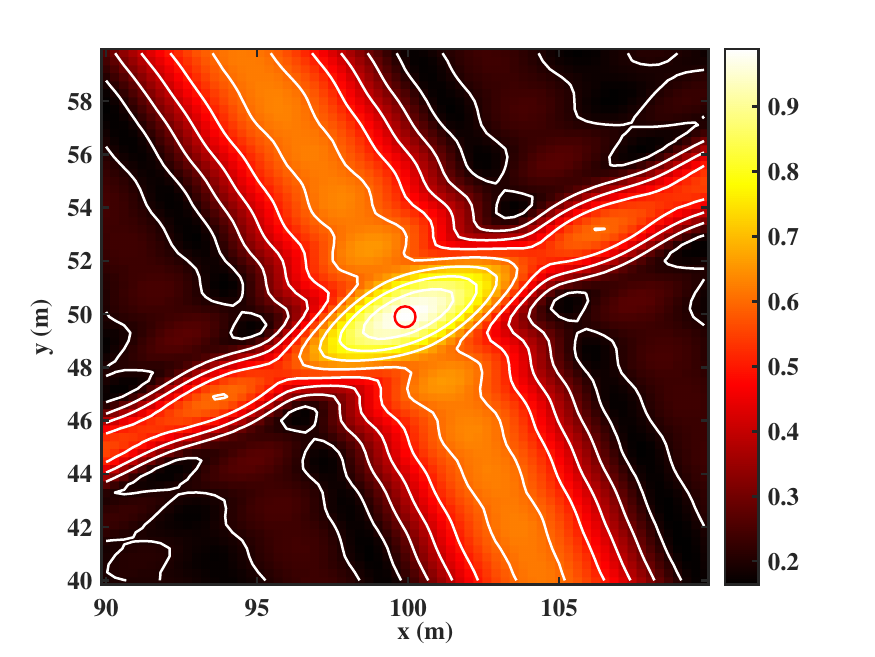}}
	\subfigure[Two MPCs] {\label{fig4.b}\includegraphics[width=.24\textwidth]{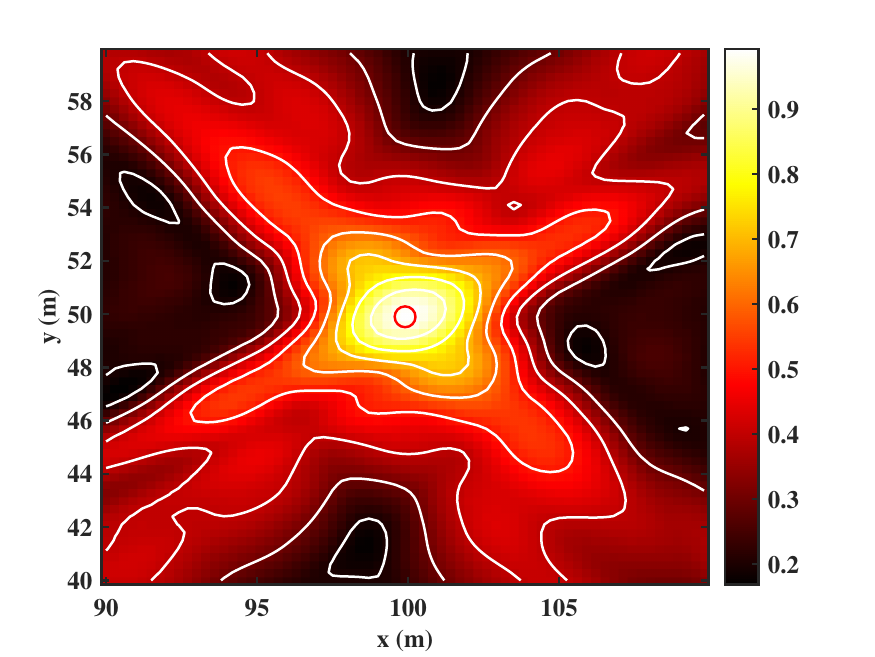}}
        \subfigure[Three MPCs]{\label{fig4.c}\includegraphics[width=.24\textwidth]{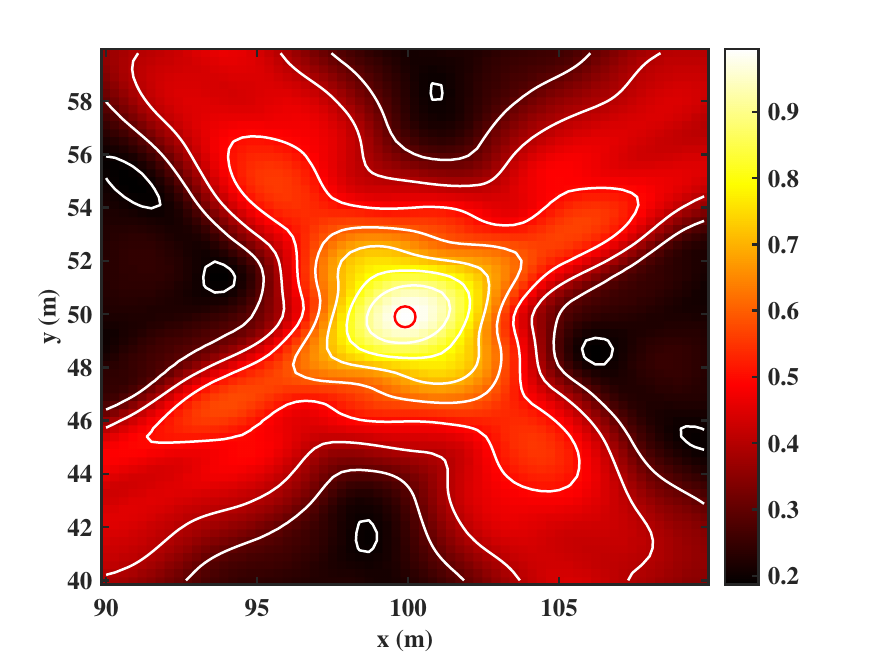}}
        \subfigure[Four MPCs] {\label{fig4.d}\includegraphics[width=.24\textwidth]{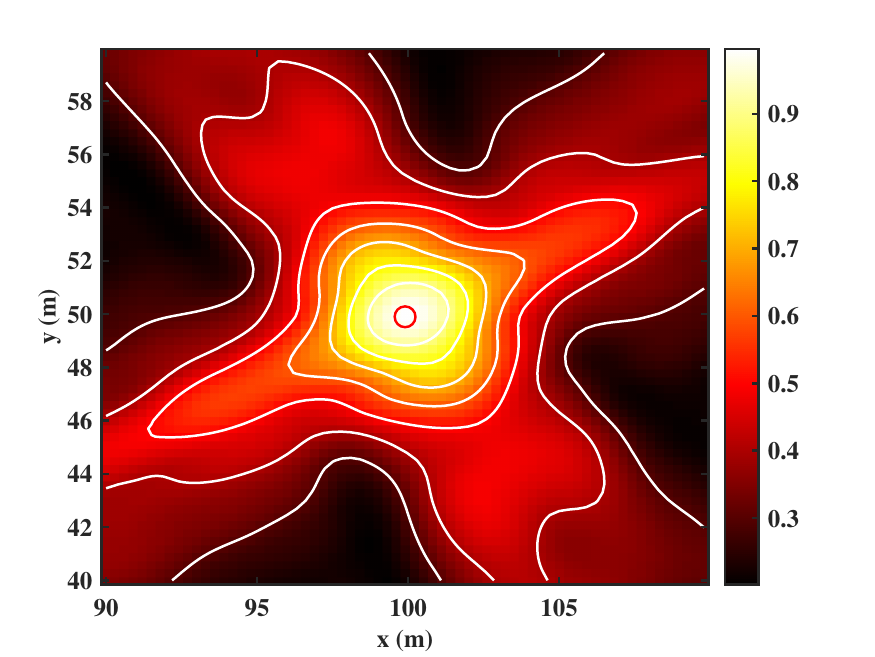}}
	\caption{Localization results with different number of MPCs in SNR = -5 dB}
	\label{fig4}
\end{figure*}

To verify the feasibility of the proposed SFMC scheme, the simulation results of localization with different number of MPCs in signal-to-noise ratio (SNR) = -5 dB are shown in Fig. \ref{fig4}, where the white lines stand for contour lines.
According to Fig. \ref{fig4}, the main lobe energy becomes concentrated as the number of MPCs increases, where the main lobe is referred to the lobe with energy higher than 0.3 in this paper. 

\begin{figure}[!htbp]
    \centering
    \includegraphics[width=0.35\textwidth]{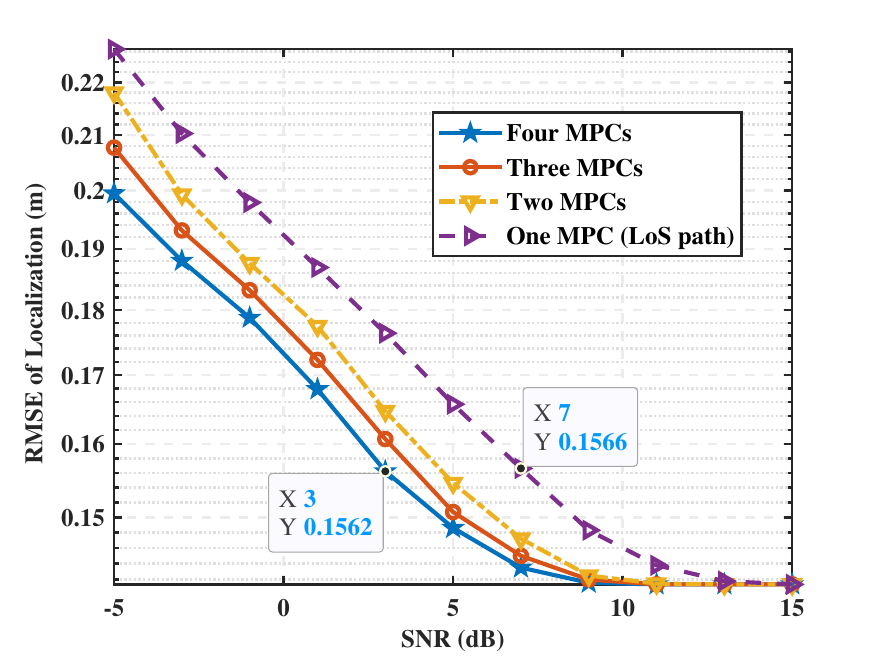}
    \caption{RMSE of localization with different number of MPCs}
    \label{fig5}
\end{figure}

Then, the RMSEs of localization with different number of MPCs are shown in Fig. \ref{fig5}.
As the number of MPCs increases, the accuracy and anti-noise performance for localization improve, demonstrating the robust and high-accuracy performance of the proposed SFMC scheme. Compared with the traditional LoS path localization scheme (i.e., with only one MPC), the SNR required to achieve the same level of localization error of 0.156 m is reduced by 4 dB. While in the high SNR regime, the RMSEs begin to level off, no matter the number of MPCs ranges from 1 to 4.

\subsubsection{Comparison of schemes}
\begin{figure}[!ht]
    \centering
    \includegraphics[width=0.35\textwidth]{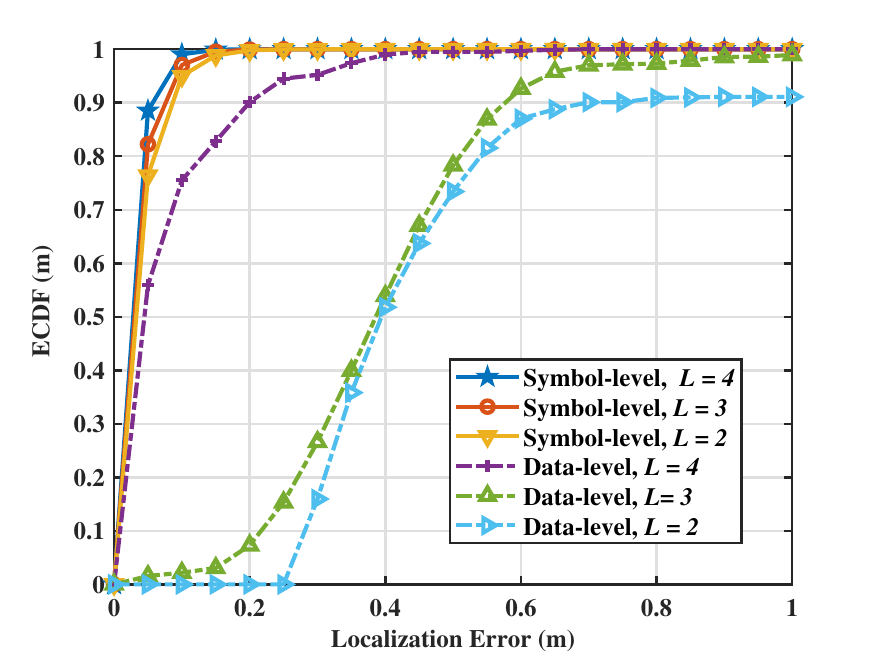}
    \caption{SFMC scheme (symbol-level) \textit{v.s.} existing scheme (data-level) }
    \label{fig6}
\end{figure}

Fig. \ref{fig6} compares the empirical cumulative distribution functions (ECDFs) of localization between the existing scheme in~\cite{gong2022multipath} and proposed SFMC scheme. As demonstrated in Fig. \ref{fig6}, the 90~\% accuracies of our scheme with four MPCs and existing scheme with four MPCs are 0.06 m and 0.2 m, respectively. The localization accuracy of the proposed SFMC scheme is around 70~\% higher than the existing scheme in~\cite{gong2022multipath}, highlighting the superiority of the proposed SFMC scheme.

\subsection{Communication Performance}
\begin{figure}
    \centering
    \includegraphics[width=0.35\textwidth]{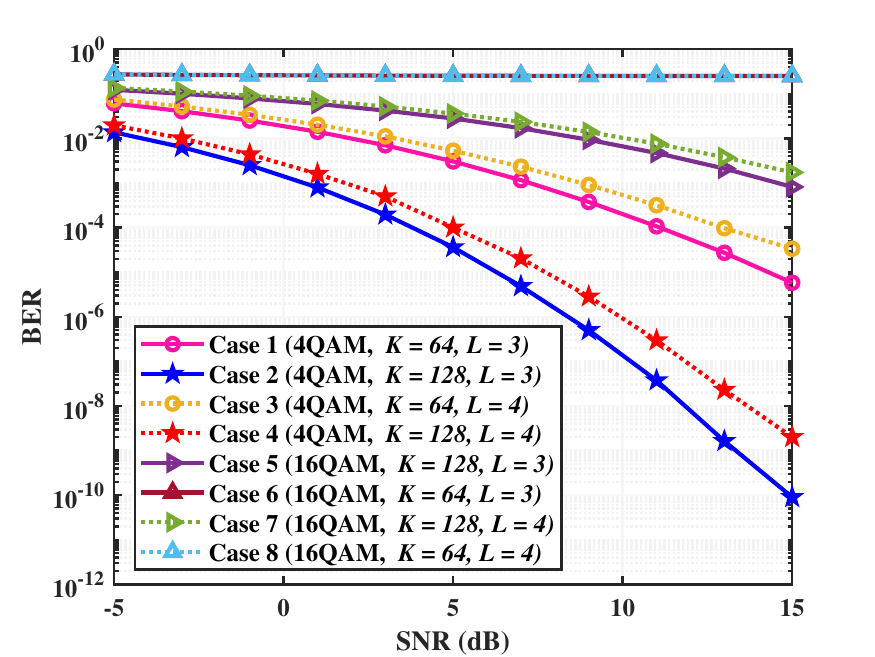}
    \caption{BER for different cases}
    \label{fig7}
\end{figure}

To investigate the impact of KRST code and MPCs on the bit error rate (BER), we have simulated different cases as shown in Fig. \ref{fig7}. {\color{blue}The transmission rates follow the order: (case 6 = case 8) $>$ (case 1 = case 3 = case 5 = case 7) $>$ (case 2 = case 4)}.

{\color{blue}Simulation results show that increasing MPCs (e.g., in case 1 and case 3) or the modulation order (e.g., in case 4 and case 7) raises the BER, reducing communication reliability. Conversely, increasing the code length (e.g., in case 1 and case 2) lowers the BER and improves reliability. For the same transmission rate (e.g., in case 1 and case 5), different combinations of modulation order and code length lead to varying performance. Thus, selecting the appropriate combination of modulation order and code length is crucial for optimizing communication performance.}

\section{Conclusion}\label{se5}
Within the concept of multi-resource cooperative ISAC, the research on MPC-aided ISAC signal processing attracts attentions. Combined with the KRST code, the MPC-aided ISAC signal processing is pioneered in this paper. To leverage the sensing information from MPC signals, a SFMC scheme is proposed, which makes use of the distribution pattern of multiple sensing information in complex space. Simulation results demonstrate that the proposed SFMC scheme achieves robust and high-accuracy localization performance compared with the existing state-of-the-art schemes.

\section*{Acknowledgment}
This work was supported in part by the Fundamental Research Funds for the Central Universities under Grant 2023RC18, in part by the National Key Research and Development Program of China under Grant 2020YFA0711302, and in part by the National Natural Science Foundation of China (NSFC) under Grant 62271081, and U21B2014.

\bibliographystyle{IEEEtran}
\bibliography{reference}

\end{document}